# Selectivity in Ligand Functionalization of Photocatalytic Metal Oxide Nanoparticles for Phase Transfer and Self-assembly Applications

Rituraj Borah[a, c], Rajeshreddy Ninakanti[a, b, c], Gert Nuyts[c, d], Hannelore Peeters[a,c], Adrián Pedrazo-Tardajos[b, c], Silvia Nuti[a, b], Christophe Vande Velde[e], Karolien De Wael[c, d], Silvia Lenaerts[a, c], Sara Bals[b, c], Sammy W. Verbruggen[a, c*]

[a] R. Borah, R. Ninakanti, H. Peeters, S. Nuti, Prof. S. Lenaerts, Prof. S. W. Verbruggen
Sustainable Energy, Air & Water Technology (DuEL), Department of Bioscience Engineering, University of Antwerp, Groenenborgerlaan 171, 2020, Antwerp, Belgium.
E-mail: Sammy.Verbruggen@uantwerpen.be

[b] R. Ninakanti, A. P. Tardajos, S. Nuti, Prof. S. Bals
Electron Microscopy for Material Science (EMAT), Department of Physics, University of Antwerp, Groenenborgerlaan 171, 2020 Antwerp, Belgium

[c] R. Borah, R. Ninakanti, G. Nuyts, H. Peeters, A. P. Tardajos, Prof. K. D. Wael, Prof. S. Lenaerts, Prof. S. Bals, Prof. S. W. Verbruggen
NANOlab Center of Excellence, University of Antwerp, Groenenborgerlaan 171, 2020 Antwerp, Belgium.

[d] G. Nuyts, Prof. K. D. Wael
Antwerp X-ray Analysis, Electrochemistry and Speciation (AXES), Department of Chemistry, University of Antwerp, Groenenborgerlaan 171, 2020, Antwerp, Belgium.

[e] Prof. C. V. Velde
Intelligence in Processes, Advanced Catalysts and Solvents (iPRACS), University of Antwerp, Groenenborgerlaan 171, 2020 Antwerpen, Belgium.

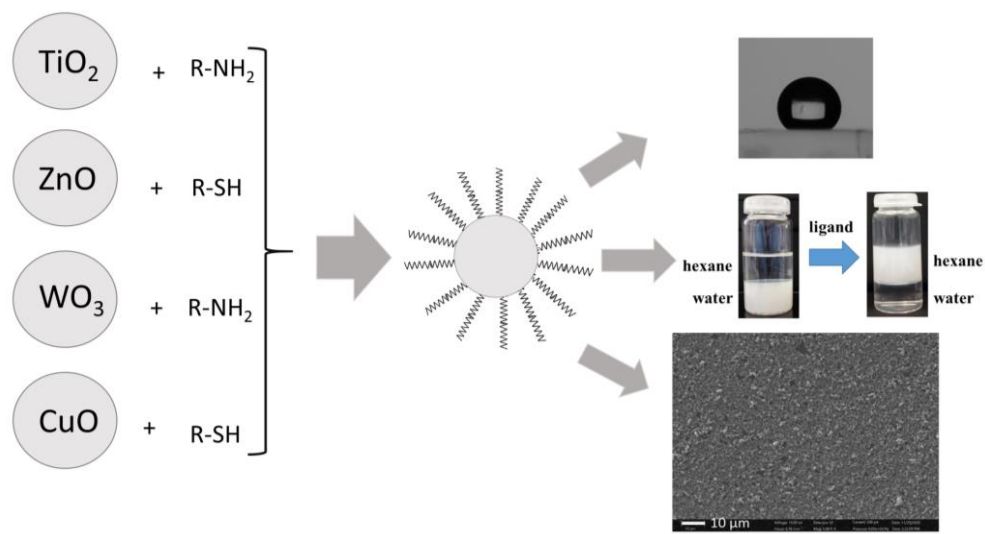

## Abstract

Functionalization of photocatalytic metal oxide nanoparticles of TiO2, ZnO, WO3 and CuO with amine-terminated (oleylamine) and thiol-terminated (1-dodecanethiol) alkyl chained ligands was studied under ambient conditions. A high selectivity was observed in the binding specificity of a ligand towards nanoparticles of these different oxides. It was observed that oleylamine binds stably to only TiO2 and WO3, while 1-dodecanethiol binds stably only to ZnO and CuO. Similarly, polar to non-polar solvent phase transfer of TiO2 and WO3 nanoparticles could be achieved by using oleylamine, but not by 1-dodecanethiol, while the contrary holds for ZnO and CuO. The surface chemistry of ligand functionalized nanoparticles was probed by ATR-FTIR spectroscopy, that enabled to elucidate the occupation of the ligands at the active sites. The photo-stability of the ligands on the nanoparticle surface was determined by the photocatalytic self-cleaning properties of the material. While TiO2 and WO3 degrade the ligands within 24 hours under both UV and visible light, ligands on ZnO and CuO remain unaffected. The gathered insights are also highly relevant from an application point of view. As an example, since the ligand functionalized nanoparticles are hydrophobic in nature, they can thus be self-assembled at the air-water interface, for obtaining nanoparticle films with demonstrated photocatalytic as well as anti-fogging properties.

## 1. Introduction

Nanoparticles of semiconductor oxides such as $TiO_2$, ZnO, $WO_3$ and CuO have gained considerable attention over the years for their proven usefulness in important applications such as photocatalysis, water splitting, $CO_2$ reduction, solar cells, super-capacitors, sensing applications, amongst others.[1] [2] [3] Many of these applications require functionalization of the nanoparticles with organic ligands with desired functionalities.[4] For instance, bio-functionalization of nanoparticles is essential for *in-vitro/in-vivo* medical applications in order to render them bio-compatible, target-specific[5] and stable.[6] Other applications of the functionalization of nanoparticles include molecular imaging[7], click-chemistry mediated conjugation[8], selective self-assembly of nanoparticles[9], ligand accelerated catalysis,[10] and many more. For nanoparticles synthesized in an aqueous medium, hydrophobic functionalization is often necessary to stably disperse the nanoparticles in a non-polar medium. Therefore, it is important to have hydrophobic functionalization strategies available that enable phase transfer of nanoparticles from an aqueous to non-polar organic phase, for follow-up reactions in this non-polar phase[11], coating of hydrophobic substrates[12], self-assembly as thin films[13], and so on.

The efficacy of the functionalization is dependent on the affinity of functional head groups towards the nanoparticles' surface.[4] Chemisorption by the formation of a covalent-like bond results in stable functionalization, while electrostatic or hydrophobic interactions result in a weaker attachment for which the surface coverage is subject to the equilibrium concentration of the ligand in the solution. The exact nature of these bonds is still an important topic of investigation even for an extensively explored case of Au and thiol.[14] [15] Recently for instance, Inkpen *et al*. challenged the widely accepted view that the Au-thiol linkage is covalent showing that the nature of the bond is mainly physisorption.[16]

Nonetheless, the strong bonding of thiols to both metal and metal-oxide nanoparticles is overall well known. Similarly, amines also bind strongly with metal as well as metal oxide nanoparticles. In heterogeneous catalysis, thiols and amines have therefore been the most explored anchoring ligands. The relative affinity of different functional groups to bind to the nanoparticles is another important aspect as it determines the success of a ligand exchange process to replace any existing ligand on the nanoparticle surface with a new ligand. In this case, the new ligands must have a stronger affinity to bind in comparison to the existing surface ligand.[17] [18]Due to the stronger affinity, oleylamine capping on Au nanoparticles can be effectively replaced by thiol ligands by simply introducing the thiol ligands to the nanoparticle colloid in a thiol-for-amine ligand exchange process.[19] Similarly, thiol-for-phosphine ligand exchange has also been achieved at room temperature due to such relative binding affinity.[20]

While functionalization of metal nanoparticles is well-explored, metal–oxide nanoparticles are far less studied in this context. In this work, we report on functionalization of semiconductor oxide nanoparticles, namely $TiO_2$, ZnO, $WO_3$ and CuO with amine (oleylamine) and thiol (1-dodecanethiol) ligands in identical chemical environments. This way the selectivity of a functional group towards a specific oxide becomes clear and is further consolidated by the fact that the efficacy of phase transfer of these nanoparticles from aqueous to non-polar (hexane or chloroform) phase is in direct correspondence with the ability of the ligand *i.e*., the head group to chemically attach on to the nanoparticles' surface. This work probes the ligand-capped nanoparticles' surface chemistry by (ATR-

) FTIR with respect to dissociative and molecularly adsorbed H$_2$O on active sites, in order to understand to what extent the different ligands occupy the active sites on the various metal oxide surfaces. Importantly, the selection of oleylamine and 1-dodecanethiol for these experiments is based on the fact that the optimum chain length is dependent on the functional group.[21][22] The alkyl chain lengths, 18-C and 12-C of oleylamine and 1-dodecanethiol, respectively, are considered optimal for phase transfer experiments. Soliwoda *et al.* showed that for thiols, 1-octadecanethiol with 18-C alkyl chains form disordered assemblies on the nanoparticle surface which results in poor phase transfer.[22] While, for amines, it takes longer 18-C chains to facilitate stable and efficient phase transfer of nanoparticles.[21] In the context of hydrophobization of nanoparticles by selective ligand functionalization, phase-transfer and air-water interfacial self-assembly, the findings from this study provide fresh insights for applications in many of these areas, of which several will be demonstrated.

## 2. Experimental section

### 2.1 Materials

The chemicals were purchased from the following suppliers and used without further purification: P25 TiO$_2$ nanoparticles (Aeroxide, size <25 nm), ZnO nanoparticles (Sigma Aldrich), WO$_3$ nanoparticles (Sigma Aldrich), CuO nanoparticles (Sigma Aldrich), oleylamine (70%, Sigma Aldrich), 1-dodecanethiol (Sigma Aldrich), absolute ethanol (Sigma Aldrich), hexane (VWR Chemicals). Deionized water with conductivity <0.1 μS cm$^{-1}$ was used. 10 mL borosilicate glass vials were used for the phase transfer.

### 2.2 Ligand adsorption experiments

Well-mixed solutions of oleylamine and 1-dodecanethiol were prepared by mixing 400 mg of each ligand in 20 ml ethanol. Colloidal solutions of TiO$_2$, ZnO, WO$_3$ and CuO Nanoparticles were prepared by mixing 50 mg of each type of nanoparticle with 10 ml of ethanol. For adsorption, 2 ml of the ligand solution were added to the nanoparticle colloids to be sonicated for 30 minutes. The colloid-ligand mixture was then left still in dark for 4 hours for adsorption. After 4 hours, the colloids were centrifuged (5000 rpm for 10 minutes) and re-washed repeatedly up to 4 cycles to remove any access unbound ligand. To obtain powders, the nanoparticles were dried in an oven at 60 °C.

### 2.3 Phase transfer experiments

A colloidal solution was prepared with 15 mg of a given type of nanoparticle (TiO$_2$/ZnO/WO$_3$/CuO) in 20 mL of water, followed by sonication for 20 minutes for uniform mixing and breaking down of agglomerates. 1 mL of this solution was mixed with 3 mL of a ligand (oleylamine or 1-dodecanethiol) solution in ethanol (200 mg in 20 mL), in a glass vial, followed by one minute of vortex mixing and 20 minutes of sonication. Now, 4 mL of pure hexane were poured onto this mixture, which remains above the water/ethanol phase as a separate phase. The glass vial was then shaken vigorously in a vortex mixture for one minute for the phase transfer of the nanoparticles to the hexane phase to occur. Upon resting the glass vial for a few minutes, the two phases separate, the water/ethanol mixture phase becomes fully transparent and the hexane phase attains the colloidal colour. The as-obtained nanoparticle in hexane colloid was centrifuged at 5000 rpm for 15 minutes to remove the excess ligands and dispersed in pure hexane. The experiment was repeated for different initial concentrations of the nanoparticle and the same oleylamine/1-dodecanethiol concentration in ethanol in order to test the efficacy.

### 2.4 Self-assembly at air-water interface and immobilization on substrate

$TiO_2$, ZnO, $WO_3$ nanoparticles were dispersed in ethanol (50 mg in 20 ml) by sonication for 15 minutes to form a stable colloid. Then, 2 ml of oleylamine in ethanol (400 mg in 20 ml) solution were added to the $TiO_2$ and $WO_3$ colloids. Similarly, 2 ml of 1-dodecanethiol in ethanol (400 mg in 20 ml) solution were added to the ZnO colloid. After addition, the colloids were subjected to 30 minutes of sonication followed by four washing steps with ethanol (500 rpm for 10 minutes). After the 4$^{th}$ step, the nanoparticles were re-dispersed in 1 ml of ethanol. 200 µl of these nanoparticles were then introduced on the walls of a beaker 80% filled with water so that the nanoparticles flow down along the wall and get trapped at the air-water interface. This step was carried on till the entire surface was covered with a nanoparticle film. In order to immobilize the self-assembled nanoparticle film on a glass or silicon surface, a dip coating method with a slow upward velocity was implemented. The films were then calcined at 450 °C for two hours, for the removal of the oleylamine or 1-dodecanethiol ligands. Alternatively, the films were subjected to UV light (Philips Cleo 25 W fluorescent light bulbs, $\lambda_{max}$ at 365 nm) for the degradation of the oleylamine via a photocatalytic route. Trials were also performed with CuO nanoparticles for which the self-assembly could not be achieved efficiently following the same procedure, as the nanoparticles tended to form agglomerates at the interface.

**2.5 Preparation of nanoparticle films on glass for contact angle measurements**

For the measurement of sessile-drop contact angles, nanoparticle films were prepared by drop-casting 200 µL of nanoparticle colloids (50 mg in 1 ml ethanol) on glass pieces previously cleaned with piranha solution. The ligand functionalized nanoparticles were also dispersed in ethanol in a similar way for drop-casting. The drop-casted films were dried in oven for one hour at 80 °C.

**2.6 Characterization**

The scanning transmission electron microscopy (STEM) images were obtained using a FEI Osiris TEM operated at 200 kV. The samples were prepared by drop casting a small drop of the colloidal solutions onto a carbon coated copper grid. The SEM characterization was done with a scanning electron microscope (SEM) equipped with secondary electron and multisegmented backscattered electron detector (EVO10, Carl Zeiss Microscopy GmbH), using an accelerating voltage of 10kV, a take-off angle of 35°. In order to characterize the organic ligand attached to nanoparticles, a Thermo Fisher Scientific Nicolet$^{TM}$ 380 FTIR spectrometer was used to record the infrared spectra in a wavenumber range of 3600 – 400 cm$^{-1}$ in both transmission and Attenuated Total Reflection (ATR) mode. The thermogravimetric analysis (TGA) measurements were performed on a TA Instruments Q5000 thermogravimetric analyzer with a sample heating rate of 10 °C min$^{-1}$ in the temperature range of 30 to 800 °C. During the photocatalytic experiments, the samples were illuminated with UVA fluorescent lamps (Philips Cleo 25 W, incident intensity of 0.87 mW cm$^{-2}$ at a distance of 7.5 cm) and visible lamps (Van-cliff minilight, 1.6 mW cm$^{-2}$ at a distance of 7.5 cm). The contact angle measurements were done with an Ossila sessile drop-based contact angle goniometer with a high-resolution camera (1920 x 1080). The droplet volume used was 4 µl. For the film thickness measurements, a Bruker DektakXT stylus profilometer was used. The measurements were done for 2 mm long sliding steps of the stylus (needle) starting from a bare surface area to the films. The powder X-ray diffraction (XRD) analysis was performed on a Bruker D8 Advance Eco (40kV, 25 mA, Cu Kα wavelength =1.5406 Å). A Low background Si sample holder was used for all XRD measurements. The scan rate was 0.04° with total time per step of 96 seconds for a 2-theta scan range of 20° to 90°.

**3. Results and discussion**

**3.1 Functionalization and phase-transfer of nanoparticles**

The FTIR spectra in Figure 1 compare the presence of the two organic ligands on the different nanoparticle powders after surface adsorption and repeated washing. The intensity of the general $v_s$(-$CH_2$-), $v_{as}$(-$CH_2$-), and $v_{as}$(-$CH_3$) stretches of the alkyl chain in the 2850-3000 cm$^{-1}$ region indicates the amount of both oleylamine and 1-dodecanethiol ligand present in the dry powder. The weaker bands at 1466 and 1377 cm$^{-1}$ arising from the asymmetric and symmetric bending vibration of the methyl group, $\delta_{as}$(-$CH_3$) and $\delta_s$(-$CH_3$), respectively, also represent the ligands. Repeated washing steps ensure that any weak physisorbed species will be removed due to mass action upon introduction of fresh solvent. Clearly, oleylamine remains bound to $TiO_2$ and $WO_3$, while 1-dodecanethiol remains on ZnO and CuO nanoparticles. It is, however, difficult to comment on the true chemical nature of the bond on the basis of the stability through washing steps. In the transmission mode, due to the higher opacity of $WO_3$, $WO_3$ sample amount had to be kept minimum yet sufficient as the spectra become too noisy for sample amounts same as the other materials.[23] The nanoparticle-ligand interaction becomes clear in the discussion of phase transfer experiments and hydrophobicity by water contact angle measurement later in this study. The broad absorption band in the 3200-3800 cm$^{-1}$ region corresponds to different bands of $H_2O$ adsorbed both dissociatively and in molecular form.[24] This broad absorption band is significantly weakened by adsorption of 1-dodecanethiol on ZnO nanoparticles. However, absorption of oleylamine does not suppress this absorption band in $TiO_2$ and $WO_3$ nanoparticles. Since water is adsorbed in many forms, it is important to differentiate those in order to derive further information, as discussed in section 3.2. Usually, the bands for amine and thiols are difficult to detect due to their relative weakness.[25][26]

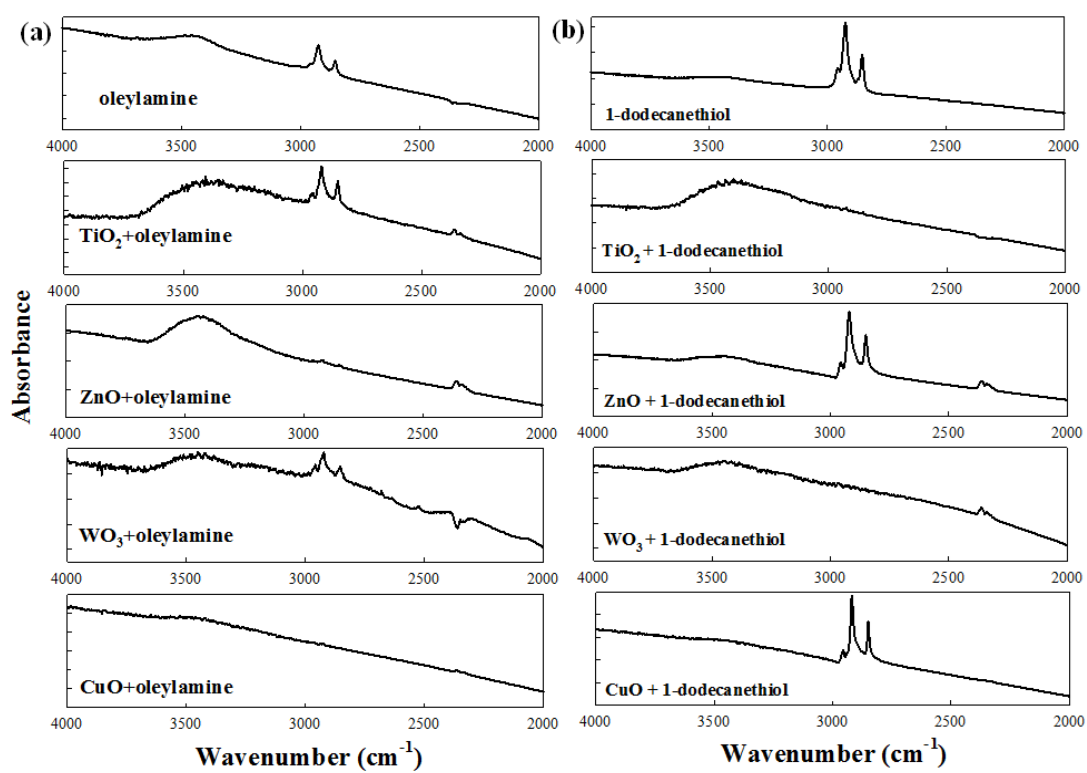

**Figure 1**. FTIR (transmission) spectra comparing $v_s$(-$CH_2$-), $v_{as}$(-$CH_2$-), and $v_{as}$(-$CH_3$) stretches of the alkyl chain of ligand functionalized nanoparticles after four washing cycles: (a) oleylamine and oleylamine + nanoparticles (b) 1-dodecanethiol and 1-dodecanethiol + nanoparticles.

Both thiol-metal and amine-metal bonding have been known to be strong interactions, while metal oxide-thiol or metal oxide-amine bonding has not been explored extensively.[27][28] In ligand-exchange studies, thiol often successfully replaces other functional groups under ambient conditions.[4][18] For instance, it has also been shown that octadecanethiol can effectively replace oleylamine on Au nanoparticle surface by ligand-exchange.[29] Generally, reports on controlled experiments comparing binding tendencies of thiols and amines on metal or metal oxide nanoparticle surfaces are scant.[30] However, considering the known strong dependency of ligand interaction on the crystal planes[31], lewis-Brønsted acid sites[32], *etc*., the selectivity of both oleylamine and 1-dodecanethiol to these four metal-oxides in Figure 1 is an expected occurrence. It has been hypothesized that ligands can coordinate as an anion to excess metal atoms on the surface to balance charge and terminate the lattice (X-type), such as oleic acid, dodecanethiol, or phosphonic acids, or as a neutral dative bond (L-type, Lewis basic), as in oleylamine, trioctylphosphine oxide, and trioctylphosphine.[33] To explain the reciprocity in the trend shown in this study, deeper insight into the atomistic interaction is required to explain the tendency of a certain functional group to a certain surface, which is a subject of ongoing theoretical/computational investigations.[34]

This selectivity of one type of functional group to a certain metal oxide observed in the adsorption experiments explains success of the phase transfer experiments of nanoparticles with a specific ligand only, as illustrated in Figure 2, and with the results of the different metal oxide-ligand phase transfer experiments summarized in Table 1. Very clearly, phase transfer with oleylamine only works for $TiO_2$ and $WO_3$, while 1-dodecanethiol only works for ZnO and CuO. Figure S1 shows the inability of the non-attaching ligands to successfully transfer nanoparticles from a water/ethanol phase to a hexane phase. The phase transfer process requires strong nanoparticle-ligand bonding in order for the nanoparticles to cross the interfacial barrier between the polar and the non-polar phases.[35,36] Apparently, the amine-ZnO or -CuO and thiol-$TiO_2$ or -$WO_3$ interactions are too weak to permanently hydrophobize the nanoparticle surfaces for the interfacial transfer. The thiol-ZnO or CuO bond are strong, resulting in smooth phase transfer and very pronounced infrared absorption bands even after repeated washing steps. It has been shown with XPS analysis that both Zn- and O-terminated nanoparticles of ZnO bind quite strongly with thiol groups and these bonds are thermally stable up to 350 °C.[37] Interestingly, on ZnO nanoparticles, an even stronger affinity of phosphonic acid functional groups as compared to thiols has been reported.[38] While certain reports on amine functionalization of ZnO and CuO nanoparticles can also be found in literature[39,40], it is clear from our experiments that the interaction is not strong enough for phase transfer. On the other hand, the strong interaction between oleylamine and $TiO_2$ or $WO_3$ easily enables phase transfer. It is important to note that a ligand that binds strongly to $TiO_2$ and $WO_3$ does not necessarily have to be an ineffective binder to ZnO or CuO surface. It is known that organic ligands with phosphonic acid functional head groups bind strongly to both $TiO_2$ and ZnO.[35,38,41,42] This consolidates the fact that the nature of the ligand-nanoparticle interaction varies case by case and the selectivity is determined by both the ligand and the nanoparticle surface chemistry. It is, however, also important to have a sufficiently high ligand concentration as the contact time needed for effective attachment is inversely proportional to this concentration.[43,44]

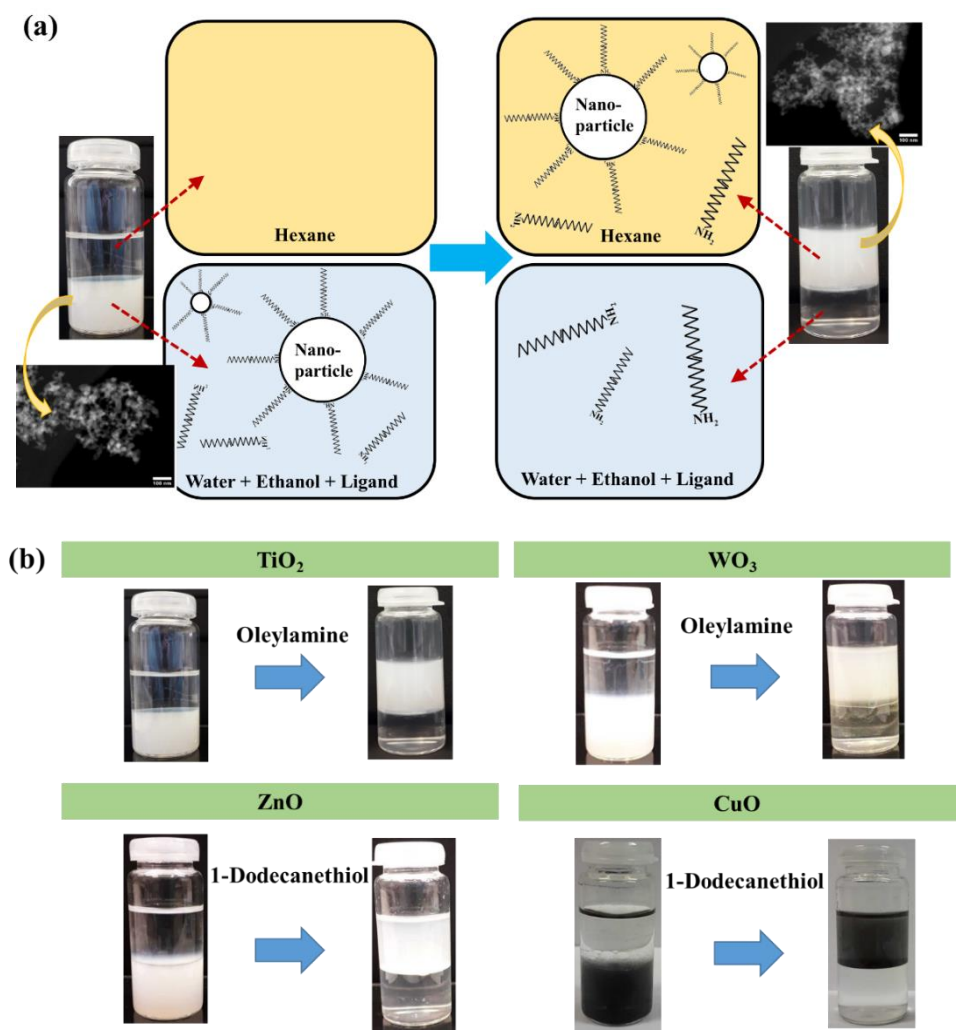

**Figure 2**. (a) Schematic explanation of the polar to non-polar phase transfer procedure of nanoparticles (inset: HAADF-STEM images of TiO$_2$ nanoparticles before and after phase transfer with 100 nm scale bar). (b) Successful phase transfer of each nanoparticle with its specific ligand.

**Table 1**. Summary of phase transfer experiments of each metal-oxide nanoparticle with oleylamine and 1-dodecanethiol.

|  | oleylamine | 1-dodecanethiol |
|---|---|---|
| TiO$_2$ | **successful** | unsuccessful |
| WO$_3$ | **successful** | unsuccessful |
| ZnO | unsuccessful | **successful** |
| CuO | unsuccessful | **successful** |

The selectivity of ligands to bind a given oxide also prompts experimentation to selectively transfer a particular oxide nanoparticle from a mixture of two. The trial experiment in Figure S2 shows that 1-dodecanethiol can selectively transfer CuO and ZnO nanoparticles from a mixture of CuO-TiO$_2$ and ZnO-TiO$_2$, respectively. However, selective phase transfer of TiO$_2$ using oleylamine was not possible possibly due to the lack of a relative binding tendency as strong as in the case of thiol. It indicates that thiol-binding is more chemically selective.

### 3.2 Surface chemistry of nanoparticle powders

Since adsorption of H$_2$O on metal oxide surfaces, especially TiO$_2$ and ZnO, is well known, comparison of pure nanoparticle powders with ligand functionalized powders with FTIR provides important insight into the ligand binding on the surface. Since the transmission mode includes the H$_2$O peaks possibly due to capillary condensation in the pores, the attenuated total reflectance (ATR) mode is more appropriate for such surface chemistry characterization. In general, ATR-FTIR is a rather simple yet efficient method for surface characterization, especially in the case of WO$_3$ that yields poor spectral information in the transmission mode due to its opacity. The broad absorption band in the FTIR spectrum of both pure TiO$_2$ and ZnO powders in the region 3000-3700 cm$^{-1}$ is commonly assigned to the stretching vibration modes of H$_2$O molecules, Figure 3(a).[45,46] In the context of surface chemistry, the band at 3695 cm$^{-1}$ is of particular interest as it represents isolated non-hydrogen-bonded OH groups on TiO$_2$, as shown in the HREELS study by Henderson, indicating availability or unavailability of active sites.[47] Similarly, the absorption band around 1637 cm$^{-1}$ is known to be from the scissoring modes of molecularly adsorbed H$_2$O, Figure 3(b). In contrast, the surfaces of WO$_3$ and CuO nanoparticles are free from adsorbed water. It has been shown by Albanese *et al.* with the help of DFT calculations that H$_2$O preferably adsorbs on WO$_3$ in its un-dissociated form due to the presence of ions at the surface that act as Lewis acid sites and this interaction per H$_2$O molecule is not strong.[48] Lu *et al.* attribute the absence of the H$_2$O peaks on commercial WO$_3$ to low surface area.[49] Similarly, a number of DFT studies has shown the possibility of both molecular and dissociative adsorption of H$_2$O on CuO surfaces[50,51], in contrast to the FTIR spectra in Figure 3, which can in fact be attributed to low surface area. Although surface adsorbed water on WO$_3$ and CuO is not apparent from the ATR-FTIR spectra, the presence of trapped moisture in the pores has been indicated in the transmission spectra in Figure 1 and TGA analysis in Figure 4. Nevertheless, comparison with respect to the surface adsorbed H$_2$O is possible only in the cases of TiO$_2$ and ZnO.

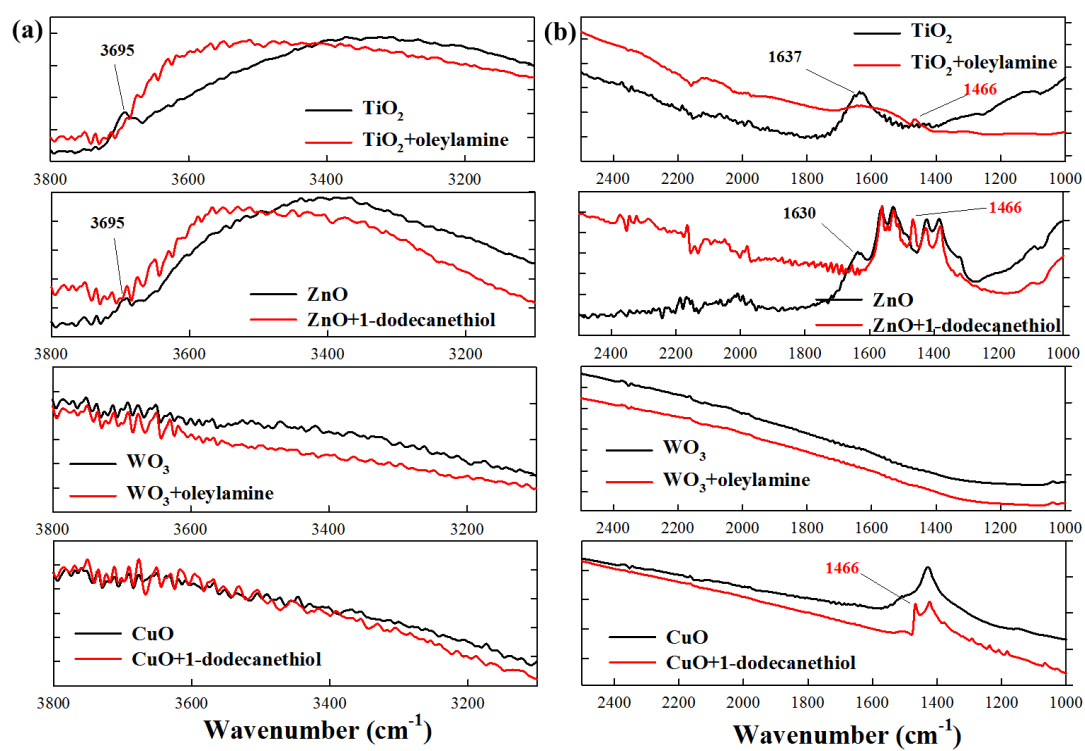

**Figure 3**. ATR-FTIR spectra of ligand functionalized nanoparticles (red) compared with their pure powders (black) in the (a) 3800-3100 cm$^{-1}$ (b) 2500-1000 cm$^{-1}$ regions in terms of H$_2$O adsorption bands.

Since H₂O adsorption on TiO$_2$ and ZnO has been studied extensively, it is useful to discuss these cases with respect to the ATR-FTIR results in order to understand the ligand-nanoparticle interaction. On the (101) surface of anatase TiO$_2$, which is dominating in this form, molecular adsorption of H$_2$O is energetically favorable in comparison to dissociative adsorption as shown by Vittadini *et al.*[52] This stability is thought to be due to the hydrogen bonding between the H atoms of the adsorbed H$_2$O molecules with bridging O atoms on TiO$_2$. On the (001) surface, dissociative adsorption is favored up to half coverage of the Ti sites beyond which molecular adsorption is more likely.[52] Physical adsorption of H$_2$O on the first layer via hydrogen bonding has also been shown to be an energetically favorable occurrence. Similarly, on the energetically favorable (10$\bar{1}$0) surface of ZnO, H$_2$O adsorbs both dissociatively and as molecules.[53][46] In the 3100-3800 cm$^{-1}$ region of broad band H$_2$O absorption in Figure 3(a), as discussed already, the shoulder at 3695 cm$^{-1}$ in both pure TiO$_2$ and ZnO distinctly signifies dissociatively adsorbed water to the surface.[47][24][54] The absence of this shoulder in the spectra of oleylamine-TiO$_2$ and 1-dodecanethiol-ZnO nanoparticles indicates unavailability of these sites for dissociative chemisorption of H$_2$O. Similarly, the bands at 1637 and 1630 cm$^{-1}$ in pure TiO$_2$ and ZnO, respectively, indicate these molecular adsorbed water bands are broadened or hidden in the ligand capped TiO$_2$ and ZnO. This also indicates occupation of these adsorption sites by the ligands. The 1466 cm$^{-1}$ band for the, $\delta_{as}$(-CH$_3$) asymmetric vibrations in both oleylamine and 1-dodecanethiol is also quite clear on ligand grafted nanoparticles. This band is particularly strong in 1-dodecanethiol-capped ZnO and CuO nanoparticles in contrast to the weak intensity in oleylamine-capped TiO$_2$ and WO$_3$ nanoparticles. As these experiments were conducted in the dark without any chance of photo-degradation, this indicates weaker oleylamine-nanoparticle interaction and reduction of surface-ligands due to mass action of washing steps.

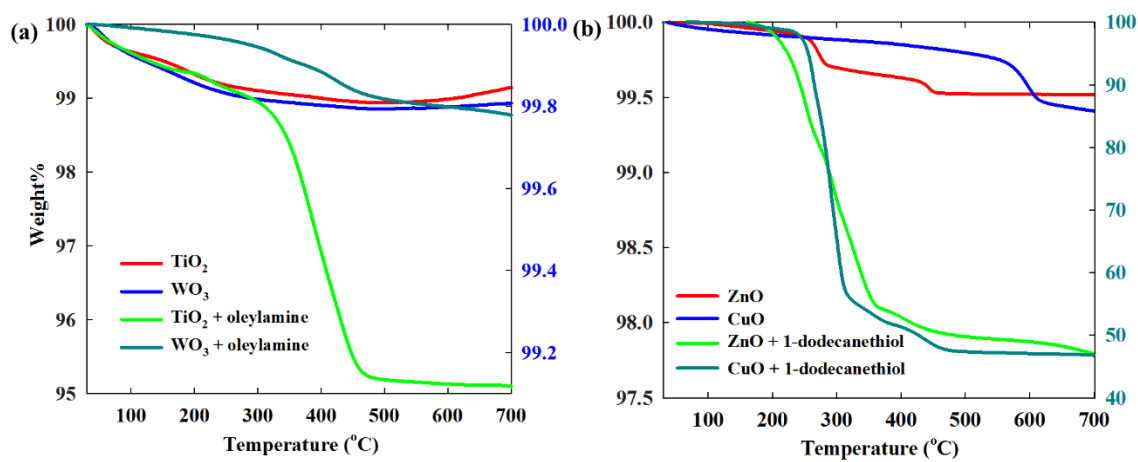

**Figure 4**. Thermogravimetric analysis (TGA) of (a) oleylamine functionalized nanoparticles (b) 1-dodecanethiol functionalized nanoparticles. Right y-axis in Figure (a): WO$_3$ powder, right y-axis in Figure (b): CuO + 1-dodecanethiol

Thermogravimetric analysis was carried out to gain further insight into the ligand-functionalization of the nanoparticles, Figure 4. 1-dodecanethiol starts to burn off at a lower temperature (~150 °C) than oleylamine (200 °C) below which both the compounds remain at their fixed initial weight, Figure S3. For the dry powders, the weight loss due to moisture loss starts at around 50 °C in all the samples. Also, for the oleylamine capped TiO$_2$ and WO$_3$, the weight loss already begins at >50 °C, hence the presence of significant trapped moisture despite ligand functionalization is evident. In the transmission FTIR spectra in Figure 1, this trapped water also contributes substantially to the 3000-3500 cm$^{-1}$ broad absorption band. In contrast, such an early weight loss is not observed in 1-dodecanethiol capped ZnO and CuO nanoparticles, meaning that moisture is present in rather low amount, which is in line with the FTIR spectra of thiol-capped ZnO and CuO nanoparticles where the $H_2O$ band intensity is significantly reduced as compared to uncapped (as oleylamine is not present) ones. This has as a possible implication that 1-dodecanethiol is able to cover ZnO and CuO nanoparticles more stably and extensively. In Figure 4, both oleylamine and 1-dodecanethiol attached on the nanoparticles start burning off at a temperature higher than their pure forms implying strong ligand-nanoparticle bonding. This elevation of the thermal decomposition temperature is again particularly well-defined for 1-dodecanethiol-capped CuO and ZnO nanoparticles due to strong binding. Importantly, for the same ligand functionalization procedure, CuO nanoparticles carry a considerably high weight % (>50%) of 1-dodecanethiol. A possible implication of this may be formation of ligand-nanoparticle networks as gel-like consistency was observed for ligand-capped CuO nanoparticles. For $TiO_2$ and $WO_3$, the transition from the moisture-loss phase to oleylamine decomposition phase is smooth. It indicates that oleylamine-nanoparticle bonding is not as strong as that in the case of 1-dodecanethiol-nanoparticle. A relatively weaker interaction between $TiO_2/WO_3$ and oleylamine has already been evident from the FTIR spectra comparisons.

### 3.3 Photo-stability of ligand capped nanoparticles

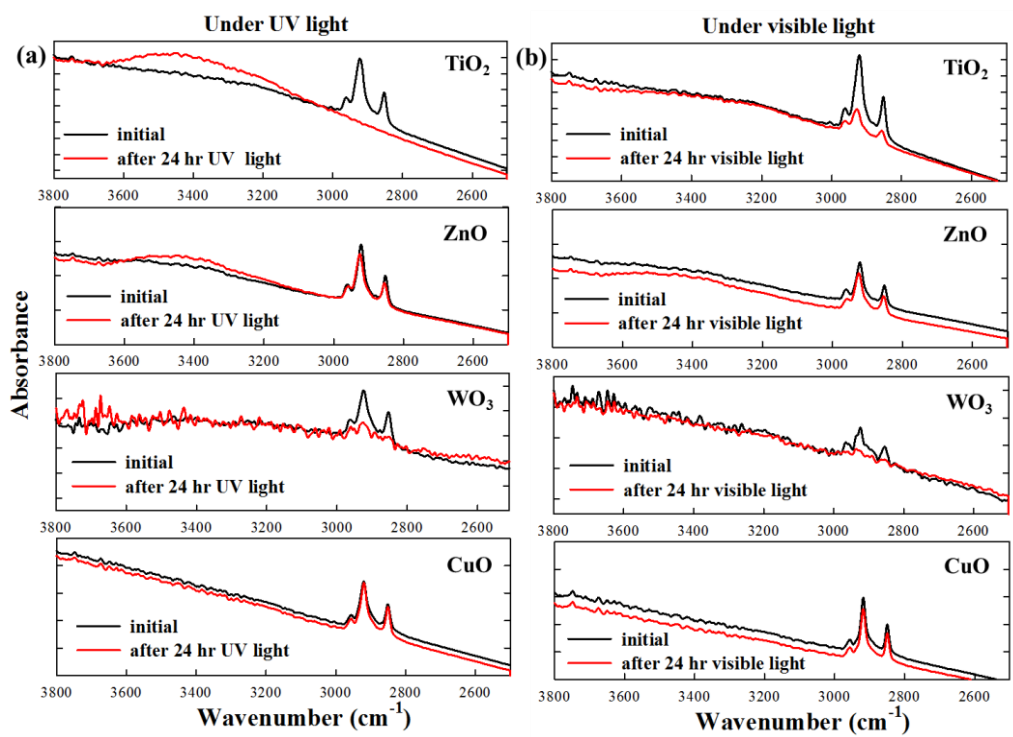

**Figure 5**. FTIR spectra ($v_s$(-$CH_2$-), $v_{as}$(-$CH_2$-), and $v_{as}$(-$CH_3$) bands in the 2800-3000 cm$^{-1}$ region) of ligand capped nanoparticle films on a Si substrate showing photo-stability of the ligands on the nanoparticles under (a) UV (b) visible light.

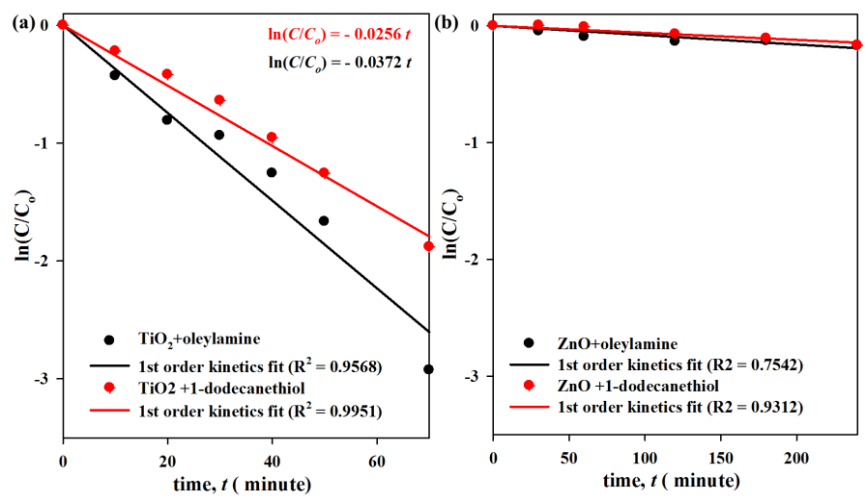

**Figure 6**. Degradation curves of oleylamine and 1-dodecanethiol on (a) $TiO_2$ and (b) ZnO films respectively fitted to a first order kinetic model. The relative concentrations are estimated from the area of the $v_s$(-$CH_2$-), $v_{as}$(-$CH_2$-), and $v_{as}$(-$CH_3$) bands in the 2800-3000 cm$^{-1}$ region.

Given that the semiconductor metal-oxide nanoparticles of $TiO_2$, ZnO, $WO_3$ and CuO are known to show photocatalytic activity, the photo-stability of the ligand-capped nanoparticles under UV and visible light is an important aspect in any possible application. Conversely, as a positive side-effect such high instability, *i.e.* fast photodegradation of the ligands, implies high self-cleaning tendencies of the nanoparticles, which in turn opens up various application opportunities. The photo-stability of the ligand capped nanoparticles was tested by monitoring the $v_s$(-$CH_2$-), $v_{as}$(-$CH_2$-), and $v_{as}$(-$CH_3$) bands of the ligands in the 2850-3000 cm$^{-1}$ region under illumination with UV and visible light, Figure 5. After 24 hours of UV illumination, $TiO_2$ completely degrades the oleylamine ligands, while $WO_3$ degrades 45%, with further degradation up to 85% after 48 hours. In contrast, ZnO and CuO nanoparticles could not degrade the ligands to any considerable extent. Similarly, also under visible light (using a white fluorescent tube lamp), both $TiO_2$ and $WO_3$ nanoparticles degrade the ligands effectively, while ZnO and CuO are unable to do the same. As expected, the photocatalytic activity of $TiO_2$ is lower under visible light with 62% degradation in 24 hours, while $WO_3$ is equally active under both UV and visible light over the 24 h time span (48% degradation in 24 hours). It is important to note that $TiO_2$ degrades the ligands under UV within an hour (Figure 6); thus, the visible light activity of $TiO_2$ is negligibly low in comparison to the UV light activity. The inactivity of ZnO, which is also a known photocatalytic material, is counterintuitive. It is reasonable to assume that the strong interaction of thiol ligands results in poisoning of the catalyst surface. Therefore, selectivity enhancements with thiol ligands are accompanied by a loss of activity.[55] To understand whether it is the inactivity of ZnO itself or a strong thiol-ZnO binding that prohibits the degradation mechanistically, degradation studies were conducted for both oleylamine and 1-dodecanethiol on $TiO_2$ and ZnO films on Si wafers, Figure 6. Instead of pre-functionalizing the nanoparticles with the ligands, oleylamine and 1-dodecanethiol was spin coated on the surface of both the $TiO_2$ and ZnO films. Clearly, $TiO_2$ degrades both oleylamine and 1-dodecanethiol rapidly within an hour. In the case of ZnO nanoparticles, the ligands remain almost unaffected even after 4 hours of illumination. The fitting of the data points to a first order kinetic model shows that the rate constant for ZnO is two orders of magnitude smaller than that for $TiO_2$. Thus, the 1-dodecanethiol capped ZnO and CuO nanoparticles are photo-stable due to their photocatalytic inactivity. On the other hand, both $TiO_2$ and $WO_3$ self-degrade the ligands effectively under both UV and visible light.

### 3.4 Wettability and self-assembly

High wettability, *i.e.* super-hydrophilicity of TiO$_2$,[56] ZnO, WO$_3$[57] and CuO[58] is known and hydrophobization of these surfaces is important in many applications. This super-hydrophilicity has mostly been attributed to a photo-induced effect explained by several surface chemistry mechanisms.[59] For more details on these mechanisms, the reader is encouraged to consult the comprehensive review by Fujishima *et al*. In addition, the switching between hydrophilic and oleophilic behavior is unique to TiO$_2$. Photo-induced hydrophilicity (PIH) has also been studied for ZnO, WO$_3$, and CuO. Like in the case of TiO$_2$, numerous studies have shown induction of hydrophilicity in ZnO and WO$_3$ films within a couple of minutes to an hour, while CuO has not been found to show any photo-induced hydrophilicity.[60][61][62] On nanoparticle films such as P25 (TiO$_2$), where the morphological irregularity may help hydrophobicity, photo-induced super-hydrophilicity can be achieved rapidly, as also verified by our experiments.[63]

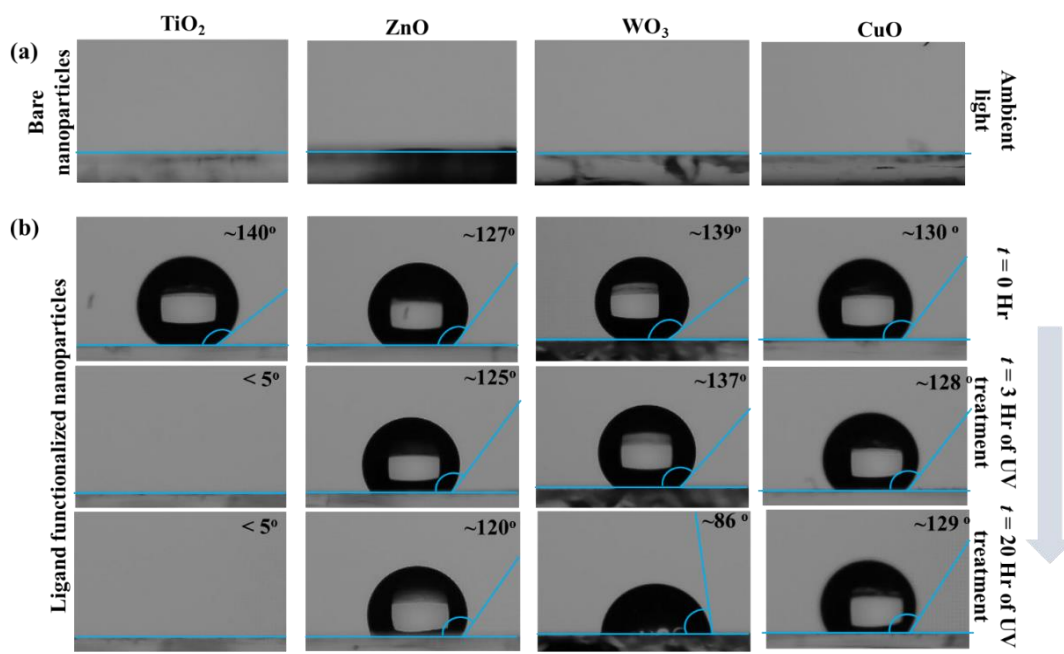

**Figure 7**. Water contact angle on drop-casted ligand capped nanoparticle films: (a) ligand free bare nanoparticles (b) hydrophobic functionalized nanoparticles, also showing the effect of UV irradiation. The contact angles are average of five measurements and the blue lines marking the angles are for representation.

So, the wettability of the ligand-capped nanoparticles of TiO$_2$, ZnO, WO$_3$ and CuO is an important aspect as the presence of the hydrophobic alkyl chains alters the surface chemistry. First, the water contact angle on drop-casted films of non-functionalized bare nanoparticle films of all four oxides were found to be close to 0°, Figure 7(a). This may be a photo-induced process as the films were prepared under ambient light. While PIH is not observed in the case of CuO, it also appears highly hydrophilic in nature. The hydrophilicity of all the different nanoparticles is confirmed by contact angle measurements which also explains easy dispersion of these nanoparticles in water to form stable colloids. In contrast, all the ligand capped nanoparticle films are strongly hydrophobic with an average water contact angle >120°, Figure 7(b). From the experiments in Figures 5 and 6, the irradiation of UV light is supposed to degrade the ligands especially on TiO$_2$ and WO$_3$ to revert the surfaces back to their hydrophilic ground state, at a rate corresponding to the photocatalytic self-cleaning rate. Thus, the super-hydrophilicity of TiO$_2$ film after 3 hours of irradiation is consistent with our photo-stability experiments showing that the ligands are completely degraded under UV within that time period. Similarly, since 24 hours of UV treatment only partially decomposed the ligands on WO$_3$ nanoparticles

(Figure 5a)), the water contact angle is still 86° after 20 hours of illumination. For both ZnO and CuO, the nanoparticle films remain hydrophobic throughout the UV treatment showing consistency with the photo-stability experiments. To confirm the effect of ligand photo-degradation on contact angle, contact angle measurements were taken after keeping the samples in dark for 24 hours, Figure S4. Importantly, while the correlation holds in this particular case, in general photocatalytic activity cannot always be directly related to photo-induced hydrophilicity as it has been shown that despite comparable photocatalytic activity of $SrTiO_3$, photo-induced hydrophilicity was not observed like in the case of $TiO_2$.[64]

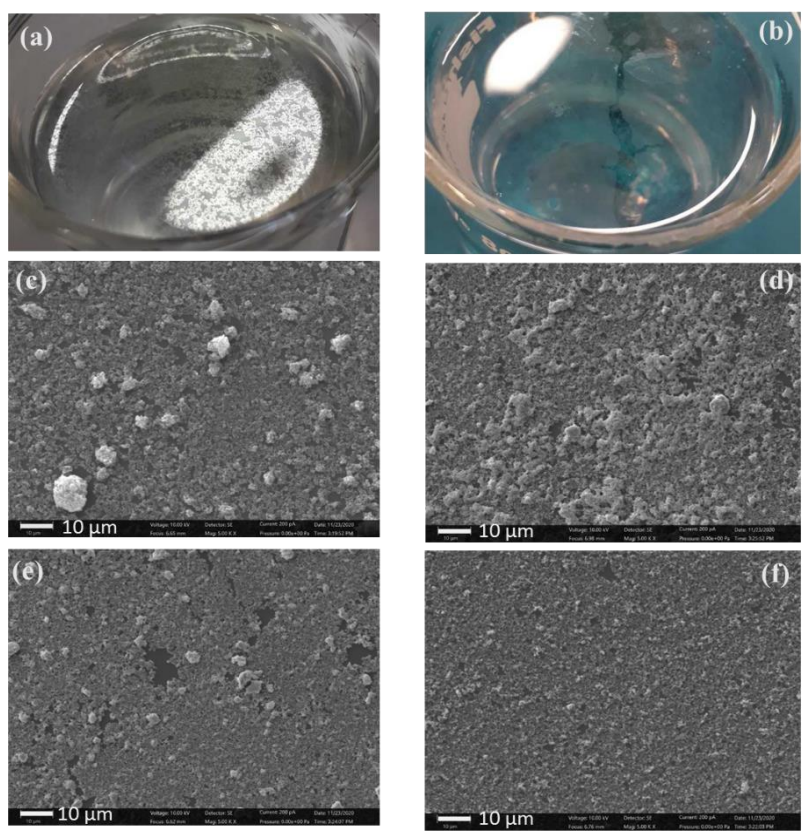

**Figure 8**. Images of assembly of ZnO (a) and $TiO_2$ (b) nanoparticles at the air-water interface. (c, d, e, f) are SEM images of self-assembled films of $TiO_2$, $WO_3$, ZnO (sample 1) and ZnO (sample 2). Sample 1: film after incomplete surface coverage; sample 2: film after complete surface coverage.

The contact angle measurement shows that the ligand capping renders the nanoparticles hydrophobic. Thus, self-assembly of these nanoparticles on a water surface is possible by trapping them at the air-water interface which is otherwise not possible, Figure S5. In Figure 8 (a), a partially covered water surface with ZnO nanoparticles is shown. As more nanoparticles are introduced, the self-assembled islands rearrange and eventually develop a continuous film. The self-assembled films obtained with ZnO nanoparticles as shown in Figure 8(e) and (f) have high uniformity. Self-assembled films of $TiO_2$ and $WO_3$ nanoparticles, Figure 8(c) and (d), however, are not as uniform due to the presence of agglomerates. While there is high surface coverage in all the cases, the uniformity of the self-assembled films is subject to the morphological uniformity of individual nanoparticles. In all these three cases of $TiO_2$, $WO_3$ and ZnO, the morphological and dimensional irregularities of the nanoparticles limit the quality of the films in terms of uniformity. Table 2 summarizes the nanoparticle average sizes as provided by the manufacturer data, crystallite size and self-assembled film thicknesses. Due to the size and morphological irregularities, estimating an average size from TEM images (Figure 2) or dynamic light scattering is not ideal. Thus, the primary crystallite size from Scherrer equation (Figure S8) by considering the peaks of significant intensity (greater than 20% of maximum intensity peaks) is provided in Table 2 as an indication of the lower limit of the particle size which matches with the manufacturer data. Film thickness profiles were acquired by stylus profilometry (Figure S6), which captures the transition as the stylus slides from the bare substrate to the film. The profilometry results capture the surface morphology only to an extent as the contact between the needle and the film leads to unavoidable displacement of some particles due to the softness of the assembled layer, hence the relatively large experimental error. Still, the measured average layer thicknesses correspond very well to the expected thickness based on a self-assembled monolayer of the corresponding particles. Even with these irregular nanoparticles, such self-assembled films can be useful in many applications. As an example, Figure 9 demonstrates the use of such a self-assembled $TiO_2$ nanoparticle film for antifogging applications.

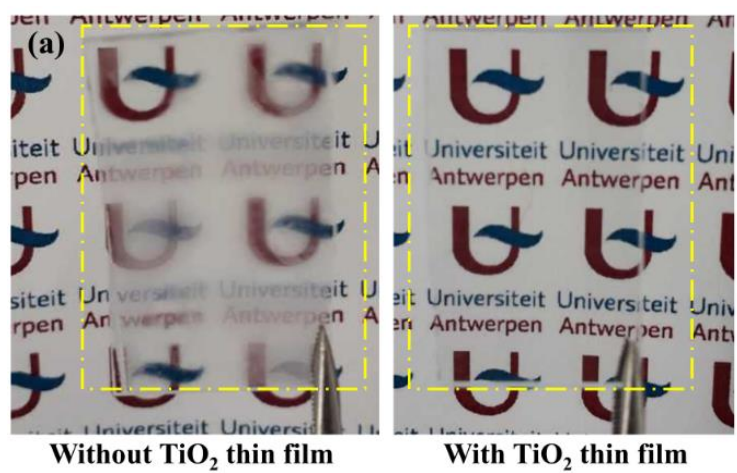

**Figure 9**. Antifogging property of a self-assembled TiO2 nanoparticle film immobilized on glass slide.

**Table 2**. Summary of nanoparticle sizes from manufacturer, crystallite sizes and self-assembled film thicknesses obtained from stylus profilometry.

| Material | Avg. Particle size (nm), manufacturer | Avg. crystallite size (nm) | Approx. Film thickness (nm) |
|---|---|---|---|
| $TiO_2$ | ≤25 nm | 25.8 ± 0.4 | 21.2 ± 4.3 |
| $WO_3$ | ≤50 nm | 48.4 ± 7.6 | 67.9 ± 13.2 |
| ZnO | ≤40 nm | 44.3 ± 5.7 | 59.3 ± 3.1 |

## 4. Conclusion

Functionalization of $TiO_2$, ZnO, $WO_3$, CuO nanoparticles with alkyl-amine (oleylamine) and alkyl-thiol (1-dodecanethiol) ligands was studied under ambient conditions. It was observed that oleylamine binds strongly to $TiO_2$ and $WO_3$, but not to ZnO and CuO. Conversely, 1-dodecanethiol strongly attaches to ZnO and CuO, but not to $TiO_2$ and $WO_3$. Polar to non-polar phase transfer experiments confirm these trends. The ligand attachment is stable through the impact of mass action in the washing steps, and confirmed by FTIR data. In the case of $TiO_2$ and ZnO, it could be shown that oleylamine and 1-dodecanethiol effectively occupy the active sites that are otherwise occupied by OH groups in the pristine powders resulting from dissociative adsorption of ambient $H_2O$. Under UV and visible light, the stability of the surface ligands on the nanoparticles was found to be dependent not only on the intrinsic photocatalytic activity of the material, but also on the ligand bond strength. While $TiO_2$ degrades oleylamine within two hours and $WO_3$ takes more than a day, ZnO and CuO appeared to be quite inactive in degrading the thiol-ligands even after 48 hours of illumination of UV or visible light. Similarly, the hydrophobicity attained by the nanoparticles can be removed by photocatalytic degradation of the amine ligands on $TiO_2$ and $WO_3$, while 1-dodecanethiol functionalized ZnO and CuO nanoparticles remain stable and hydrophobic under UV and visible light illumination over an extended period of time (>48 hours). The functionalized nanoparticles can be easily self-assembled at the air-water interface to obtain nanoparticle films for various applications, such as antifogging layers as a demonstrated proof-of-principle in this work. While the ligand selectivity of different metal oxide nanoparticles and the corresponding phase transfer processes shown in this work is an important aspect relevant to various application scenarios, the demonstration of the use of these hydrophobic functionalized nanoparticles for self-assembled films is potentially valuable to future research on photocatalytic surfaces, photo-electrochemical and photovoltaic applications. Especially in view of the ongoing research on hybrid multifunctional nanoparticles, self-assembly methods are important for obtaining films in a nonintrusive way.

**Conflicts of interest**

There is no conflict of interest to declare.

**Acknowledgment**

R.B. and S.W.V. acknowledge financial support from the University of Antwerp Special Research Fund (BOF) for a DOCPRO4 doctoral scholarship. S.B. and A.P.-T. acknowledge financial support from the European Commission under the Horizon 2020 Program by means of the grant agreement No. 731019 EUSMI and the ERC Consolidator Grant No. 815128 REALNANO**.****References** the stability of the surface ligands on the nanoparticles was found to be dependent not only on the intrinsic photocatalytic activity of the material, but also on the ligand bond strength. While $TiO_2$ degrades oleylamine within two hours and $WO_3$ takes more than a day, ZnO and CuO appeared to be quite inactive in degrading the thiol-ligands even after 48 hours of illumination of UV or visible light. Similarly, the hydrophobicity attained by the nanoparticles can be removed by photocatalytic degradation of the amine ligands on $TiO_2$ and $WO_3$, while 1-dodecanethiol functionalized ZnO and CuO nanoparticles remain stable and hydrophobic under UV and visible light illumination over an extended period of time (>48 hours). The functionalized nanoparticles can be easily self-assembled at the air-water interface to obtain nanoparticle films for various applications, such as antifogging layers as a demonstrated proof-of-principle in this work. While the ligand selectivity of different metal oxide nanoparticles and the corresponding phase transfer processes shown in this work is an important aspect relevant to various application scenarios, the demonstration of the use of these hydrophobic functionalized nanoparticles for self-assembled films is potentially valuable to future research on photocatalytic surfaces, photo-electrochemical and photovoltaic applications. Especially in view of the ongoing research on hybrid multifunctional nanoparticles, self-assembly methods are important for obtaining films in a nonintrusive way.

**Conflicts of interest**

There is no conflict of interest to declare.

**Acknowledgment**

R.B. and S.W.V. acknowledge financial support from the University of Antwerp Special Research Fund (BOF) for a DOCPRO4 doctoral scholarship. S.B. and A.P.-T. acknowledge financial support from the European Commission under the Horizon 2020 Program by means of the grant agreement No. 731019 EUSMI and the ERC Consolidator Grant No. 815128 REALNANO**.**

**References**

(1) Huang, Z.-F.; Song, J.; Pan, L.; Zhang, X.; Wang, L.; Zou, J.-J. Tungsten Oxides for Photocatalysis, Electrochemistry, and Phototherapy. *Advanced Materials* **2015**, *27* (36), 5309–5327. https://doi.org/10.1002/adma.201501217.
(2) Medhi, R.; Marquez, M. D.; Lee, T. R. Visible-Light-Active Doped Metal Oxide Nanoparticles: Review of Their Synthesis, Properties, and Applications. *ACS Appl. Nano Mater.* **2020**, *3* (7), 6156–6185. https://doi.org/10.1021/acsanm.0c01035.
(3) Verbruggen, S. W. TiO2 Photocatalysis for the Degradation of Pollutants in Gas Phase: From Morphological Design to Plasmonic Enhancement. *Journal of Photochemistry and Photobiology C: Photochemistry Reviews* **2015**, *24*, 64–82. https://doi.org/10.1016/j.jphotochemrev.2015.07.001.
(4) Sperling, R. A.; Parak, W. J. Surface Modification, Functionalization and Bioconjugation of Colloidal Inorganic Nanoparticles. *Philosophical Transactions of the Royal Society A: Mathematical, Physical and Engineering Sciences* **2010**, *368* (1915), 1333–1383. https://doi.org/10.1098/rsta.2009.0273.
(5) Pollinger, K.; Hennig, R.; Ohlmann, A.; Fuchshofer, R.; Wenzel, R.; Breunig, M.; Tessmar, J.; Tamm, E. R.; Goepferich, A. Ligand-Functionalized Nanoparticles Target Endothelial Cells in Retinal Capillaries after Systemic Application. *Proceedings of the National Academy of Sciences* **2013**, *110* (15), 6115–6120. https://doi.org/10.1073/pnas.1220281110.
(6) Conde, J.; Dias, J. T.; Grazú, V.; Moros, M.; Baptista, P. V.; de la Fuente, J. M. Revisiting 30 Years of Biofunctionalization and Surface Chemistry of Inorganic Nanoparticles for Nanomedicine. *Frontiers in Chemistry* **2014**, *2*.
(7) Thiruppathi, R.; Mishra, S.; Ganapathy, M.; Padmanabhan, P.; Gulyás, B. Nanoparticle Functionalization and Its Potentials for Molecular Imaging. *Advanced Science* **2017**, *4* (3), 1600279. https://doi.org/10.1002/advs.201600279.